\documentclass[useAMS,usenatbib,aastex]{mn2e}

\usepackage{graphicx}
\usepackage{amssymb,amsmath}
\usepackage{natbib}

\newcommand{\ain}{a_{\rm in}}
\newcommand{\aout}{a_{\rm out}}
\newcommand{\argsh}{{\rm argsh}}
\newcommand{\sech}{{\rm sech}}
\newcommand{\elie}{{\mathbf E}}
\newcommand{\elid}{{\mathbf D}}
\newcommand{\elih}{{\mathbf H}}

\newcommand{\elik}{{\mathbf K}}
\newcommand{\elipi}{{\mathbf \Pi}}

\newcommand{\elitun}{{\mathbf T}_1}
\newcommand{\elitdeux}{{\mathbf T}_2}

\newcommand{\greenfinh}{{\cal G}^{\rm in.}}
\newcommand{\greenf}{{\cal G}}
\newcommand{\greenfa}{{\cal G}_{\rm app.}}
\newcommand{\tk}{\tilde{k}}

\newcommand{\araa}{ARA\&A}
\newcommand{\aap}{A\&A}
\newcommand{\apjs}{ApJS}
\newcommand{\apj}{ApJ}
\newcommand{\aj}{AJ}
\newcommand{\mnras}{MNRAS}

\newcommand{\slfrac}[2]{\left.#1\middle/#2\right.}
\begin{document}

\title[The  potential of discs from a ``mean Green function'']{The  potential of discs from a ``mean Green function''}
\author[A. Trova, J.-M. Hur\'e and F. Hersant]
{A. Trova$^{1,2}$\thanks{E-mail:audrey.trova@obs.u-bordeaux1.fr},
J.-M. Hur\'e$^{1,2}$\thanks{E-mail:jean-marc.hure@obs.u-bordeaux1.fr} and
F. Hersant$^{1,2}$\thanks{E-mail:franck.hersant@obs.u-bordeaux1.fr} \\
$^{1}$Univ. Bordeaux, LAB, UMR 5804, F-33270 Floirac, France\\
$^{2}$CNRS, LAB, UMR 5804, F-33270 Floirac, France}

\date{Accepted ???. Received ???}

\pagerange{\pageref{firstpage}--\pageref{lastpage}} \pubyear{???}

\maketitle

\label{firstpage}

\begin{abstract}

By using various properties of the complete elliptic integrals, we have derived
an alternative expression for the gravitational potential of axially symmetric bodies, which
is free of singular kernel in contrast with the classical form. This is mainly a radial integral of the
local surface density weighted by a regular ``mean Green function'' which
depends explicitly on the body's vertical thickness. Rigorously, this result stands for a
wide variety of configurations, as soon as the density structure is vertically
homogeneous. Nevertheless, the sensitivity to vertical stratification | the Gaussian profile has been considered | appears weak provided that the surface density is conserved. For bodies with small aspect ratio (i.e. geometrically thin discs), a first-order
Taylor expansion furnishes an excellent approximation for this mean Green
function, the absolute error being of the fourth order in the aspect ratio.
This formula  is therefore well suited to studying the structure of self-gravitating discs and
rings in the spirit of the ``standard model of thin discs'' where the vertical
structure is often ignored ,but it remains accurate for discs and tori of finite thickness.
This approximation which perfectly saves the properties of Newton's law
everywhere (in particular at large separations), is also very useful for
dynamical studies where the body is just a source of gravity acting on external
test particles.  
\end{abstract}

\begin{keywords}
Accretion, accretion discs | Gravitation | Methods: analytical | Methods: numerical
\end{keywords}

\section{Introduction}
\label{sec:intro}

The capability to calculate properly the gravitational potential inside
celestial bodies is a longstanding challenge in Astrophysics, mainly due to the
difficulty to manage the hyperbolic divergence $\slfrac{1}{|\vec{r}-\vec{r}'|}$
(a major feature of Newton's law). The presence of strong inhomogeneities of
density and complex geometries or shapes seem less problematic in comparison.
For a few special configurations (not always realistic), the potential/density
pair is simple and analytical \citep{mestel63,binneytremaine87}. However, for
most realistic matter distributions, a numerical computation of the
gravitational potential has to be performed, either from the integral form when
accuracy is required \citep{av83,hachisu86,stonenorman92,baruteaumasset08}, or
from the Poisson equation, when computational time is the limiting factor
\citep{bodo92,fbd04}. While most theoretical developments allow to reach any
degree of accuracy, their numerical implementation are usually limited by
errors and uncertainties of various origins.  A striking illustration is the
infinite series representation of the Green function in terms of Legendre
polynomials which, although exact, is known to have a (very) low convergence
rate inside sources \citep[see e.g.][]{kellogg29,durand64}; so, its
implementation is often disappointing in practice
\citep[]{clement74,hachisu86}. Compact integral expressions, despite singularities, may
then be preferred \citep{ct99,hurepierens05}.

The gravitational potential of discs, starting with galactic discs, has
received much attention in the last decades \citep{mestel63,binneytremaine87}.
Even for very simple and academic configurations, new analytic solutions and
approximations continue to be produced
\citep{cg02,schulz09,vl10,vl11,hureetal11,ba11,schulz11}, either for
establishing theoretical references, or for analysing observations. The problem
is far from being solved. Searching for new formulae or techniques aiming at
improving the description (in terms of computation time, accuracy, series convergence,
singularity avoidance, etc.) of gravitational potentials and forces of
celestial bodies is therefore still of relevance today. 

In a series of papers \citep{hurepierens05,hh07,hpelatp07,hureetal08,hh11}, we
have mainly investigated flat, axially symmetric discs where the surface
density is a power-law of the equatorial radius, with a special attention to
edge effects. In this paper, we consider, as in \cite{ph05}, systems with
finite thickness, and present a new decomposition of the gravitational Green
kernel which permits to express the potential as a sum of two regular
integrals. These two integrals have interesting properties and their
expression can give rise to very accurate approximations in the limit of
geometrically thin discs, i.e. when the vertical thickness $h$ of the disc is
small enough.

This paper is organized as follows. In Section \ref{sec:theory}, we recall the
general expression for the potential valid under axial symmetry and define the
associated ``mean Green function'' to be calculated. In Section
\ref{sec:integrable_singularity}, we demonstrate that this mean function | an
improper integral | can be rewritten as the sum of a regular, line integral
over the boundary of the system and of an analytical function, also regular.
This enables to rewrite the expression for the potential in a different, more
tractable form which contains a surface integral and a line integral. This is
the aim of Section \ref{sec:generalexp}. In Section \ref{sec:lr}, we briefly
verify that the new expression for the potential automatically fulfills the
right conditions at infinity. The question of the reduction of the remaining
surface integral is addressed in Section \ref{sec:reduction}. In Section
\ref{sec:approx}, we build an approximation for the mean Green function from a
first-order Taylor expansion (systems with small aspect ratios). We check the
accuracy of the approximation and associated potential in Section
\ref{sec:numex} through a typical toroidal configuration. In Section \ref{sec:inhomogeneous}, we show that stratification effects remain weak inside and outside the source, as soon as the total surface density is conserved. The last Section is
devoted to a conclusion. A few useful formulae and demonstrations are found in
Appendix.

\section{The gravitational potential under axial symmetry. Theoretical grounds}
\label{sec:theory}
\begin{figure}
\centering
\includegraphics[width=8.3cm]{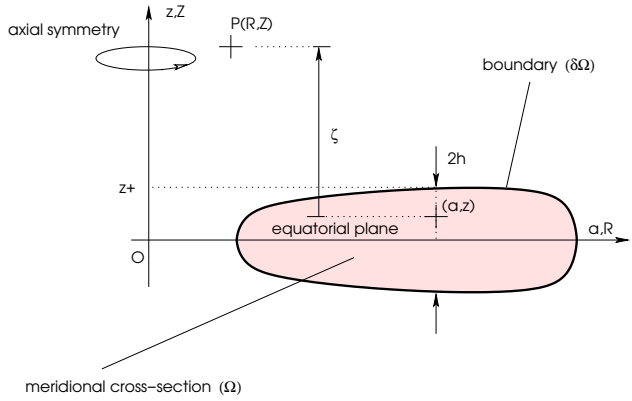}
\caption{Typical configuration for the gravitating, axially symmetric body (finite size and mass, and local total thickness $2h$) and notations associated with the cylindrical coordinate system.}
\label{fig:axibody.eps}
\end{figure}

\begin{figure}
\centering
\includegraphics[width=7.8cm, bb=73 46 709 529, clip=]{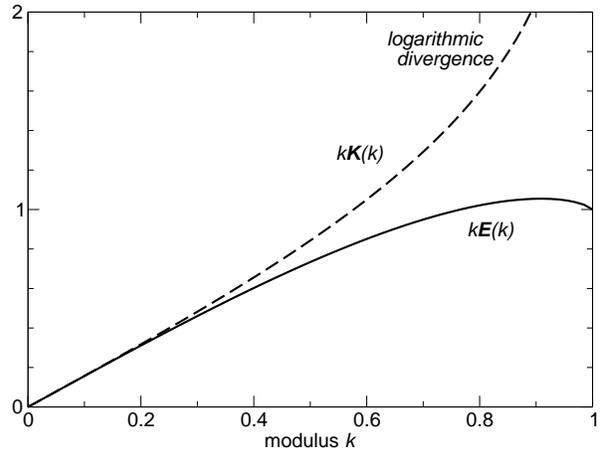}
\caption{Variation of $k\elik(k)$ and $k \elie(k)$ with the modulus $k$. The first function logarithmically diverges as $k \rightarrow 1$, while the second function is bounded (and takes its maximal value for $k \approx 0.91$).}
\label{fig:ek.eps}
\end{figure}

We consider an axially symmetric disc defined by its meridional cross-section
$(\Omega)$ and mass density $\rho$, as depicted in Fig. \ref{fig:axibody.eps}.
The gravitational potential $\psi$ of this body is given, in cylindrical
coordinates $(R,Z)$, by the surface integral \citep[e.g.][]{durand64}:

\begin{equation}
\psi(R,Z)= - 2G \iint_{(\Omega)}{\sqrt{\frac{a}{R}} \rho (a,z) k \elik(k) da dz},
\label{eq:psit}
\end{equation}
where $(a,z)$ refers to points inside $(\Omega)$ belonging to the source,
$\elik(k)$ is the complete elliptical integral of the first kind (see the
Appendix \ref{app:a} for its definition),

\begin{equation}
k = \frac{2\sqrt{aR}}{\sqrt{(a+R)^2+\zeta^2}} \in [0,1],
\label{eq:kz}
\end{equation}
is the modulus, and $\zeta = Z-z$ is the altitude difference. The divergence of
the term $\slfrac{1}{|\vec{r}-\vec{r}'|}$ when $\vec{r} \rightarrow \vec{r}'$
initially present in Newton's law still exists here through the function
$\elik(k)$ which logarithmically diverges\footnote{Precisely, we have
\citep{gradryz65}: 
\begin{equation}
\elik(k) \underset{k\rightarrow 1}{\sim} \ln 4 - \ln \sqrt{1-k^2}
\end{equation}
\label{note:lnkprim}} as $k \rightarrow 1$ (corresponding to
$a=R$ and $z=Z$). This singularity is clearly visible in Fig. \ref{fig:ek.eps}
where we have plotted the function $k \elik(k)$ versus $k$. Although $\elik(k)
\rightarrow \infty$ as $k \rightarrow 1$, the integral of the potential is
finite \citep{kellogg29,durand64}. Outside the material domain $(\Omega)$,
Eq.(\ref{eq:psit}) can be easily evaluated numerically as the modulus $k$ is
always less than unity. But difficulties begin as one approaches the boundary
($\partial \Omega$). Inside the source, the direct computation of the integral
by standard numerical techniques cannot lead to accurate potential values due
to the improper integral (at the location where $k$ reaches unity). The separate
treatment of the logarithmic divergence (see note \ref{note:lnkprim}) gives
good results in terms of precision \citep{ansorg03,hure05,ba11}, but this
approach is not very tractable. For these reasons, the Green function
is generally replaced by its series expansion in Legendre polynomials. This
effectively removes the divergence but generates new problems in practice
\citep[low convergence rate of the alternate series, truncations, etc.; see
e.g.][]{clement74,hachisu86}.

The way to perform the double integral in Eq.(\ref{eq:psit}) depends more or
less on the mass density distribution and body's shape via the equation for its
boundary $(\partial \Omega)$. If this equation can be defined by two bijections
of the form $z_\pm(a)$ (i.e. $z_+$ above the equatorial plane and $z_-$ below),
then we can perform the integral over $z$ first, followed by the integral over
$a$.  This is the most natural approach to treat geometrically thin discs and
rings \citep[e.g.][]{ss73,pringle81}, and this is the case we will consider in
the following. The body is then regarded as a collection of {\it infinitely
thin cylinders}. Assuming that the mass density does not depend upon $z$ but
only varies with the radius $a$, Eq.(\ref{eq:psit}) can be written in the form:

\begin{equation}
\psi(R,Z)= - G \int_a{ \greenf\left(R,Z;a,z_\pm(a)\right) \Sigma (a) da},
\label{eq:psit2}
\end{equation}
where
\begin{equation}
\Sigma (a) = \int_z{\rho(a)dz}
\end{equation}
is the total surface density in the disc, and
\begin{equation}
\label{eq:meang}
\greenf \equiv \slfrac{\int_{z} 2 \sqrt{\frac{a}{R}} k \elik(k) dz}{\int_{z}{dz}}
\end{equation}

plays the role of a ``mean Green function''. Actually, $\elik(k)$ rises around
$a=R$ and $\zeta =0$, while it has small amplitude elsewhere ($\elik(k)
\rightarrow \frac{\pi}{2}$ for small $k$). Due to the integration process,
${\cal G}$ is expected to be a regular function,  peaking around $a=R$ (see
below). It seems that {\it this integral is not known in closed-form}, but only
through an alternate series whose convergence rate is unfortunately
low \citep{durand64}\footnote{Actually, in contrast with the two components of
the gravitational acceleration due to a {\it thin cylinder}, the potential is
apparently available only through the series \citep{durand64}.}.

For certain configurations where $z_\pm(a)$ is not bijective, it may be more
convenient to integrate over $a$ first, and then over $z$. This procedure is
equivalent to considering the body as infinitely flat discs piled up along the
$z$-direction\footnote{The formula for homogeneous, infinitely thin disc is
given in \cite{durand64} \citep[see
also][]{lassblitzer83,fukushima10,tresaco11}.}. Finally, it is worth mentioning that the double integral over the meridional cross-section can be
converted, under certain conditions, into a line integral over the boundary
$(\partial \Omega)$ through the curl theorem which writes:

\begin{equation}
\int_{(\Omega)}{(\partial_{a}N - \partial_{z}M) dadz} = \int_{(\partial \Omega)}{(Mda+Ndz)},
\end{equation}
under axial symmetry. Here, this requires the existence of two cylindrical functions $M(a,z)$ and $N(a,z)$ such that:
\begin{equation}
-2\sqrt{\frac{a}{R}}\rho k \elik(k) = \partial_{a}N - \partial_{z}M.
\label{eq:fromcurl}
\end{equation}

This is not guaranteed in the general case where density gradients are present.
However, \cite{ansorg03} have solved this question in the fully homogeneous case,
i.e. when $\rho$ is a constant.

\section{Bypassing the kernel singularity: the case of vertically homogeneous systems}
\label{sec:integrable_singularity}

The absence of any closed-form for $\greenf$ is problematic for most
theoretical and numerical applications. We can however rewrite this mean Green
kernel in another, more convenient form, as follows. By using various
relationships involving the complete elliptic integrals of the first, second
and third kind ($\elik$, $\elie$ and $\elipi$ respectively; see the Appendix
\ref{app:a} for their definition), we have established by direct calculus the
following two formulae, valid for any integers $n$ and $p$:

\begin{flalign}
\nonumber
\partial_\zeta \zeta^n k^p \elik(k)  =  \frac{\zeta^{n-1} k^p}{m^2} & \left\{ \left[(n-p+1)m^2+(p-1)k^2\right] \elik(k)  \right.\\
&\left. +  \left(k^2-m^2\right) \frac{\elie(k)}{{k'}^2}  \right\},
\label{eq:dzetankpk}
\end{flalign}
and
\begin{flalign}
\nonumber
\partial_\zeta \zeta^n k^p \elipi(m,k)& = \frac{\zeta^{n-1} k^p}{m^2} \left\{ \left[(n-p)m^2+(p-1)k^2\right] \elipi(m,k)\right.\\
& \left. +  k^2 \frac{\elie(k)}{{k'}^2}  \right\}.
\label{eq:dzetankppi}
\end{flalign}
where
\begin{equation} 
m=\frac{2\sqrt{aR}}{a+R},
\end{equation}
is the characteristic or parameter of $\elipi$, and
\begin{equation} 
k'=\sqrt{1-k^2}
\end{equation}
is the modulus complementary of $k$. Note that $m \rightarrow k$ as
$\zeta~\rightarrow~0$ and $0 \le k \le m \le 1$. If we now combine together
Eqs.(\ref{eq:dzetankpk}) and (\ref{eq:dzetankppi}), with weights $1$ and
${m'}^2=1-m^2$ respectively, we get the general formula:

\begin{flalign}
& \partial_\zeta  \zeta^n k^p \elik(k) - {m'}^2 \partial_\zeta \zeta^n k^p \elipi(m,k) =  \frac{\zeta^{n-1} k^p}{m^2} \\
\nonumber
&\times \left\{ \left[(n-p+1)m^2+(p-1)k^2\right] \elik(k) \right.\\
\nonumber
& \qquad \left. -  {m'}^2 \left[(n-p)m^2+(p-1)k^2\right] \elipi(m,k) -  m^2 \elie(k) \right\}.
\end{flalign}
The remarkable point is that, for $p=1$ and $n=p$, the complete elliptic of the third kind disappears in the right-hand-side, and we get:
\begin{equation}
\partial_\zeta \zeta k \elik(k) - {m'}^2 \partial_\zeta \zeta k \elipi(k)  = k \elik(k) - k \elie(k).
\end{equation}
As $m$ does not dependent on the relative altitude $\zeta$, we can write this relation as:
\begin{equation}
\partial_\zeta \left\{ \zeta \left[ k \elik(k) - {m'}^2  k \elipi(k) \right] \right\}  = k \elik(k) - k \elie(k).
\label{eq:integ_sing}
\end{equation}
Finally, given $d\zeta=-dz$, it follows that\footnote{\cite{hd12} make another use of this relationship.}:
\begin{equation}
\int{k\elik(k) dz} = \int{k\elie(k) dz} - \zeta \elih(m,k)
\label{eq:integ_sing_2}
\end{equation}
where $\elih$ is defined by
\begin{equation}
\elih(m,k) =  k \left[ \elik(k) - {m'}^2 \elipi(m,k) \right].
\end{equation}

We recognize in the left-hand-side of Eq.(\ref{eq:integ_sing_2}) the expression
coming into the definition of our ``mean Green function'', in
Eq.(\ref{eq:meang}). This new form is very interesting. Actually, the first
term in the right-hand-side of Eq.(\ref{eq:integ_sing_2}) is fully regular
since $\elie(k) \in [\frac{\pi}{2},1]$. The function $k \elie(k)$ is displayed
versus $k$ in Fig. \ref{fig:ek.eps} and it is to be compared with the initial
kernel $k \elik(k)$. The second term is also fully regular and bounded.
The divergence of $\elih(m,k)$ when $k$ and $m$ both approach unity
is cancelled out by the presence of the two vanishing factors $m'^2$ and
$\zeta$. The $\elih$-function is plotted versus $m$ and $k/m$ in Appendix
\ref{app:b}. We conclude that Eq.(\ref{eq:integ_sing_2}) has a three
satisfactory properties: i) it no longer contains any singular integrand, ii) it
ensures that the effect of initial singularity is automatically accounted for,
and iii) it fully saves the Newtonian properties of the potential and
associated forces.

\section{A general expression for the Newtonian potential of vertically homogeneous bodies}
\label{sec:generalexp}

By inserting Eq.(\ref{eq:integ_sing_2}) in Eq.(\ref{eq:meang}), we find that
the mean Green function associated with vertically homogeneous systems writes:
\begin{flalign}
\nonumber
\greenf\left(R,Z;a,z_\pm\right) & = \sqrt{\frac{a}{R}} \frac{1}{h} \left[\int_{z_-}^{z_+}{k\elie(k) dz} \right. \\
& - \left. \zeta_+ \elih(m,k_+) + \zeta_- \elih(m,k_-)  \right],
\label{eq:integ_sing_3}
\end{flalign}
where, from Eq.(\ref{eq:kz}):
\begin{equation}
\label{eq:kplusmoins}
k_\pm=\frac{2\sqrt{aR}}{\sqrt{(a+R)^2+\zeta_\pm^2}},
\end{equation}

and $\zeta_\pm=Z - z_\pm$ is eventually a function of $a$ (sign $+$ being
associated with the top boundary, and sign $-$ is for the bottom). 
Since $k_\pm = 1$ only when $\zeta_\pm=0$, the $\zeta \elih$ terms in
Eq.(\ref{eq:integ_sing_3}) are always finite. 
It means that the integration of the mean Green function over the
radius $a$ | see Eq.(\ref{eq:psit2}) | can be performed numerically without
much difficulty as soon as $z_\pm(a)$ is known. In this integration process,
some care must be taken though. Actually, when regarded as a function of $a$,
$\elih(m,k)$ is not derivable at $R=a$. The jump in the derivative can, however,
be determined (see Sect. \ref{sec:numex} and Appendix \ref{app:c}). Depending
on $(R,Z)$, this jump is:

\begin{itemize}
\item $0$, outside the body,
\item $\pi$, on the boundary $(\partial \Omega)$,
\item $2\pi$, inside the domain $(\Omega)$.
\end{itemize}
As a result, $\greenf$ is peaked at $a=R$, and the maximum value $\greenf_{\rm max.}$ is:
\begin{flalign}
\label{eq:maxgreenf}
\greenf_{\rm max.} & \equiv \greenf\left(R,Z;R,z_\pm\right)\\
\nonumber
 & = \frac{1}{h(R)} \left[ \int_{z_-}^{z_+}{k\elie(k) dz}  - \zeta_+ k_+ \elik(k_+) + \zeta_- k_- \elik(k_-)  \right].
\end{flalign}

Note that, for a flat body, $z_\pm = 0$ and so one recovers the well-known expression:
\begin{equation}
\greenf\left(R,Z;a,0\right) = 2 \sqrt{\frac{a}{R}} k\elik(k).
\label{eq:gflat}
\end{equation}

The general expression for the gravitational potential is then found from Eqs. (\ref{eq:psit2}) and (\ref{eq:integ_sing_3}), namely:
\begin{flalign}
\label{eq:psit3}
\psi(R,Z) & = - G \int_a \Sigma(a) \frac{1}{h}\sqrt{\frac{a}{R}} da \int_{z_-}^{z_+} k\elie(k) dz  \\
\nonumber
& +G \int_a{ \Sigma(a) \sqrt{\frac{a}{R}} \frac{1}{h} \left[\zeta_+ \elih(m,k_+) - \zeta_- \elih(m,k_-) \right]da}.
\end{flalign}

We recall that this expression is exact and valid for any point $(R,Z)$ of
space, for any surface density profile $\Sigma(a)$ (no vertical density
gradients), and semi-thickness $h(a)$. It contains a surface integral over
$(\Omega)$ and a line integral over $(\partial \Omega)$ through $z_\pm(a)$.

If the body is symmetric with respect to the equatorial plane, we have $z_+ =
-z_-= h$. Unfortunately, Eq.(\ref{eq:integ_sing_3}) does not simplify more
since $k_+$ and $k_-$ remain different, except for $Z=0$. This case
corresponds to the potential at the midplane. With $\zeta_+ = - \zeta_-=-h$,
the midplane mean Green function writes:

\begin{equation}
\greenf\left(R,0;a,h\right) = 2 \sqrt{\frac{a}{R}}\left[ \frac{1}{h} \int_0^h{k\elie(k) dz}+ \elih(m,k_\pm)   \right]
\label{eq:integ_sing_midplane}
\end{equation}
and
\begin{flalign}
\greenf_{\rm max.} = 2 \left[ \frac{1}{h(R)} \int_0^h{k\elie(k) dz}  + k_\pm \elik(k_\pm)  \right].
\end{flalign}
\section{Long-range properties}
\label{sec:lr}

We can check that the long-range properties of the potential defined by
Eq.(\ref{eq:psit3}) are correct. Actually, far enough from the system, the
potential is expected to tend towards the potential due to a central condensation, or
\begin{equation}
\lim_{r \rightarrow \infty} r \psi = -GM,
\end{equation}
where $r=\sqrt{R^2+Z^2}$ is the spherical radius, and 
\begin{equation}
M=2\pi \int_a{ a \Sigma(a) da}\\
\end{equation}
is the total mass. We see that $k \approx 2\sqrt{aR}/r  \rightarrow 0$ as
$r~\rightarrow~\infty$, and then $\elie(k) \rightarrow \slfrac{\pi}{2}$. It
means that the long-range behavior is ensured by the first term in the
right-hand-side of Eq.(\ref{eq:psit3}) containing the complete elliptic
integral of the second kind $\elie$, whereas the contribution from the
$\elih$-function is a higher-order correction (see Appendix \ref{app:d} for a
more detailed analysis).

\section{Full reduction to a one-dimensional integral ?}
\label{sec:reduction}

The important question we have tried to clarify concerns the possibility to
convert the remaining double integral in the right-hand-side of
Eq.(\ref{eq:psit3}) into a line integral (by performing the integral over $z$
or $\zeta$). Actually, the existence of a compact expression for

\begin{equation}
\int{k \elie(k) d\zeta}
\label{eq:keofkdzeta}
\end{equation}
would be helpful, since the potential $\psi$ of a tri-dimensional, vertically
homogeneous and axially symmetric system would be given by a one-dimensional
integral. This is the reason why we have explored different paths, but
unsuccessfully. We have for instance re-written $k \elie(k)$ in the form of
an infinite series over $k$, or re-considered $\elie(k)$ as an integral over
$\phi$, followed by term-by-term integration as done in \cite{cvijovic94}.
However, none of these two approaches leads, apparently, to a closed-form in
the general case. Equation (\ref{eq:keofkdzeta}) can also be converted into
various equivalent forms, like:

\begin{flalign}
\label{eq:variousforms}
\int{k \elie(k) d\zeta} = \pm \frac{4aR}{a+R} \int{ \frac{\elie(k)dk}{k\sqrt{m^2-k^2}}},
\end{flalign}
where $\pm = |\zeta|/\zeta$. This integral over $k$ is known only for $m=1$. In
particular, the indefinite form has been derived by Dieckmann (2011; private
communication\footnote{See \tt http://pi.physik.uni-bonn.de/\~{}dieckman/}),
but it involves hypergeometric series which are not easy to manipulate. The
definite form\footnote{The definite integral can take various equivalent form,
like:
\label{not:sech}
\begin{flalign}
\label{eq:variousforms2}
\int_\infty^{\zeta_+}{k \elie(k) d\zeta} & = 2\sqrt{aR} \int_\infty^{u_+}{ \elie\left(m \, \sech(u)\right) du}\\
\nonumber
& = 2\sqrt{aR} \int_0^{v_+}{ \frac{\elie(m v) dv}{vv'}}\\
\nonumber
& = \frac{2\sqrt{aR}}{k_+} \int_0^{1}{ \frac{\elie(tk_+)dt}{t\sqrt{1-\frac{k_+^2}{m^2}t^2}}} 
\end{flalign}
where
\begin{equation}
\zeta = (a+R)\sinh u,
\end{equation}
\begin{equation}
v = \sech(u),
\end{equation}
and
\begin{equation}
t = \frac{k}{k_+},
\end{equation}
which can unfortunately not be reduced more, because formulae linking complete
elliptic integrals with different moduli, for instance $\elie(\alpha x)$ and
$\elie(x)$ for any real $\alpha \in [0,1]$, are missing (Landen and Gauss
transformations are useless here).} is known only when the integral bounds are
$0$ and $1$ \citep{cvijovic94,Cvijovic1999169,Prudnikov88} which corresponds to
$\zeta$ in the range $]\pm \infty,0]$. This is a very special situation: $a$
and $R$ are obviously different in general, and so the case $m=1$ is too
restrictive. Finally, following \cite{ansorg03}, we have also searched for a
two-component vector $(M',N')$ whose curl would yield the function
$\sqrt{\frac{a}{R}}k\elie(k)$ (see Sect. \ref{sec:theory}), but we failed to find a
simple and convincing answer. This question remains therefore open.

\section{Application to geometrically thin discs and rings: a useful and accurate approximation}
\label{sec:approx}

The main assumption made up to now is the absence of vertical stratification
for the mass density. It is true that, except for very academic (i.e.
unrealistic) configurations, this may a priori not apply to astrophysical
problems | however, as we show in Section \ref{sec:inhomogeneous}, the sensitivity to stratification remains weak. Possible domains of interest concern certain polytropic fluids \citep[e.g.][]{hachisu86,ansorg03}, and geometrically thin discs and rings \citep{pringle81}. For thin discs actually, radial gradients are often neglected over vertical gradients \citep{pringle81}, at least far from edges.
In this context, a vertically averaged structure with a uniform density along the
$z$-axis is commonly assumed \citep{ss73,cd90,dubrulle92}. Their total thickness $2h(a) \equiv z_+-z_-$ is everywhere small compared to the radius $a$, which enables many developments. Regarding the present problem, it means that
$k_+$ and $k_-$ are always close to each other. We can, in this case, produce
an approximation for the mean Green function (and thus for the potential)
by estimating Eq.(\ref{eq:keofkdzeta}).
There are various ways to address the problem. In order to get an approximation
valid in the whole physical space (and not only in the disc vicinity as often
considered), it is appropriate to work with the mean modulus $\tk$ defined by:

\begin{equation}
\tk = \frac{k_{+}+k_{-}}{2}.
\label{eq:meank}
\end{equation} 

\begin{figure}
\centering
\includegraphics[width=7.8cm, bb=73 46 709 529, clip=]{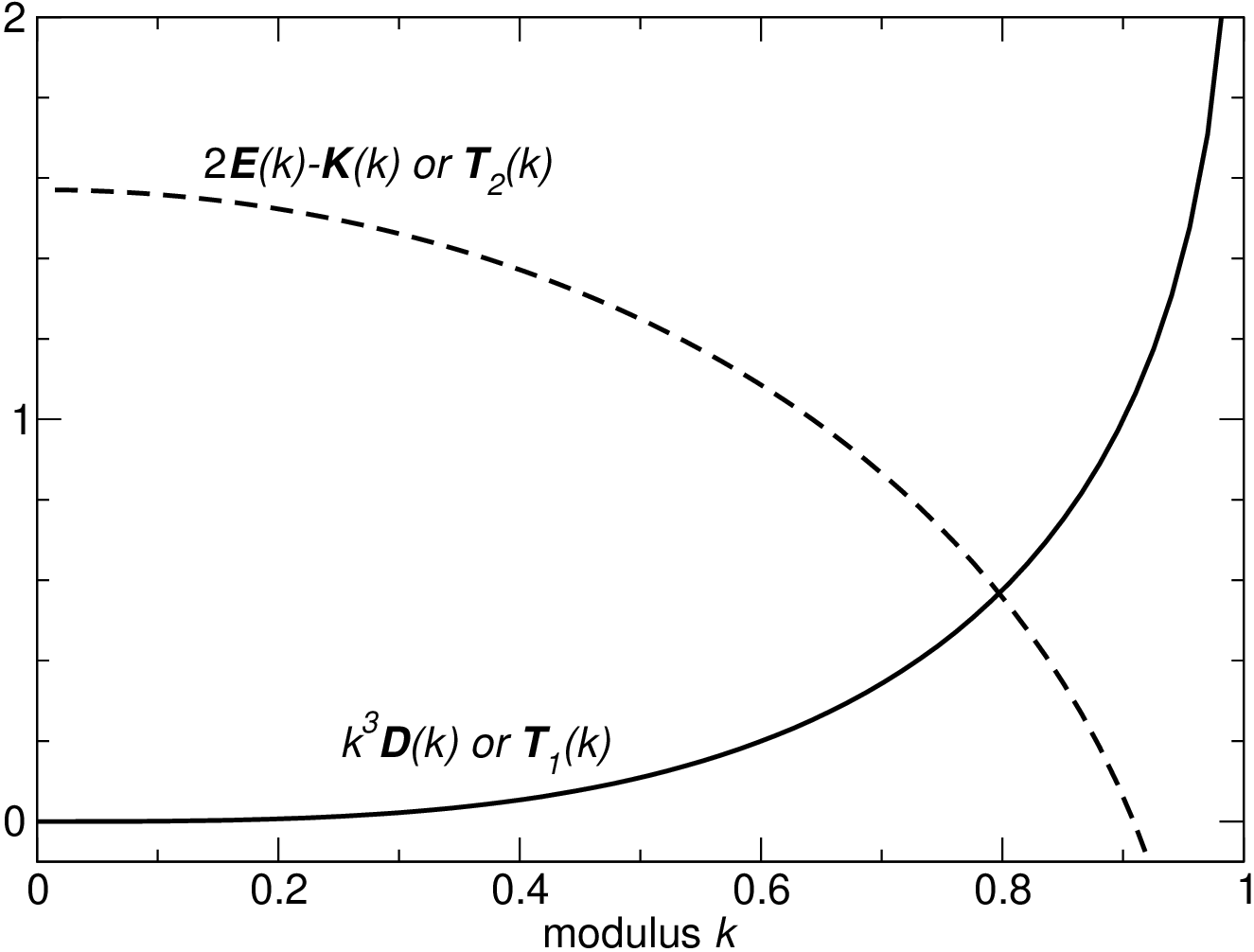}
\caption{Variation of $\elitun(k)$ and $\elitdeux(k)$ with the modulus $k$. These two functions logarithmically diverge as $k \rightarrow 1$.}
\label{fig:t1t2.eps}
\end{figure}

In particular, $\tk$ never reaches unity\footnote{This is true except when
$z_+=z_-$ and $Z=0$. This case is met when the potential is requested in the
equatorial plane, at radii where the thickness is zero. This is precisely the
case of ``edges'' where the mass density $\rho$ is also expected to vanish. So,
we can conclude that $\tk <1$ for realistic configurations.}, preventing any
eventual divergence of the elliptic integrals. If we now perform a Taylor
expansion of $k\elie(k)$ around $\tk$, we obtain, at first order:

\begin{equation}
k\elie(k) \approx \tk\elie(\tk)+ \left. \frac{d}{dk} k\elie(k) \right|_{\tk}(k-\tk),
\end{equation}
and then (see the appendix \ref{app:a} for the derivative):
\begin{equation}
k\elie(k) \approx \tk\elie(\tk)+\left[2\elie(\tk)-\elik(\tk)\right](k-\tk),
\end{equation}
which is always bounded since $\tk < 1$. The truncation produces an error of
the order of $(k-\tk)^2$. We can now estimate the integral of $k\elie(k)$ with
respect to $z$ or $\zeta$. After re-arranging terms, we can write this integral
as:
\begin{equation}
\label{eq:intapprox}
\int_{\zeta} k\elie(k) d\zeta \approx \elitun(\tilde{k}) \zeta + \elitdeux(\tilde{k}) \int_{\zeta} k d\zeta
\end{equation}
where we have defined:
\begin{equation}
\elitun(k)= k \left[\elik(k)-\elie(k)\right] \equiv k^3 \elid(k),
\end{equation}
and
\begin{equation}
\elitdeux(k)= 2\elie(k)-\elik(k) \equiv \elie(k)- k^2 \elid(k).
\end{equation}

where $\elid(k)$ is another common complete elliptic integral (see the
Appendix \ref{app:a}). These two functions are plotted versus $k$ in Fig.
\ref{fig:t1t2.eps}. Note that $\elitun$ and $\elitdeux$ logarithmically diverge
as $k \rightarrow 1$. This is not a problem since these functions are
always invoked with a modulus less than one. Besides, the integration of $k$ with respect to $\zeta$ is easily found analytically:
\begin{flalign}
\label{eq:intkdz}
\int_\zeta kd\zeta & = \int_\zeta{\frac{2\sqrt{aR}}{\sqrt{(a+R)^2+\zeta^2}} d\zeta}\\
\nonumber
& = 2\sqrt{aR} \, \argsh \left( \frac{\zeta}{a+R} \right).
\end{flalign}

By replacing in Eq.(\ref{eq:integ_sing_3}) the vertical integral by
Eq.(\ref{eq:intapprox}), its approximation, we find the ``approximate mean
Green function'':
\begin{flalign}
\greenfa & \left(R,Z;a,z_\pm\right)  = 2 \sqrt{\frac{a}{R}}  \left\{ \elitun(\tilde{k}) \right. \\
\nonumber
& -  \elitdeux(\tilde{k}) \frac{\sqrt{aR}}{h} \left[  \argsh \left( \frac{\zeta_+}{a+R} \right) - \argsh \left( \frac{\zeta_-}{a+R} \right) \right] \\
\nonumber
& \left.  - \frac{\zeta_+}{2h} \elih(m,k_+) + \frac{\zeta_-}{2h} \elih(m,k_-)  \right\}.
\label{eq:approxmeang}
\end{flalign}
where $\tk$ is given by Eq.(\ref{eq:meank}). The error is expected to be of the order of $\int_z{(k-\tk)^2dz}$. The approximation for the potential is given by Eq.(\ref{eq:psit2})
where $\greenf$ is to just be replaced by $\greenfa$, or:
\begin{equation}
\label{eq:psit_approx}
\psi(R,Z) \approx -G \int_a{ \Sigma(a)  \greenfa\left(R,Z;a,z_\pm\right) da}.
\end{equation}

As this approximation works in the whole physical space (no hypothesis has been
made upon $R$ and $Z$), it can be used not only to model the
internal structure of self-gravitating discs and rings, but also for dynamical
studies where the system is basically a source of gravity acting on external,
moving particles \citep[e.g.][]{subrkaras05,tresaco11}. Equation
(\ref{eq:psit_approx}) is also attractive from a numerical point of view as it
involves a one dimensional integral, with finite integrand.

The long-range properties of the potential are automatically conserved here as,
in deriving this approximation, we have made no hypothesis about the relative
distance to the system. When $r \rightarrow \infty$, we have $\tk \sim
2\sqrt{aR}/r \rightarrow 0$,  and so $2\elitdeux(\tk)  \sim  \pi$ while
$4\elitun(\tk)  \sim  \pi \tk^2$ (the $\elih$-function brings a similar,
second-order correction; see the Appendix \ref{app:e}). The main contribution
in the potential then comes from $\elitdeux(\tk)$:
\begin{flalign}
\psi(R,Z) &\approx -G\int{\Sigma(a) \sqrt{\frac{a}{R}} (- \pi \tk) (-2h) da } \\
\nonumber
& \sim  -4 \pi G  \frac{1}{r} \int{\Sigma(a) a da }\\
\nonumber
&  \sim  - \frac{GM}{r}.
\end{flalign}

\begin{figure}
\includegraphics[width=8.3cm]{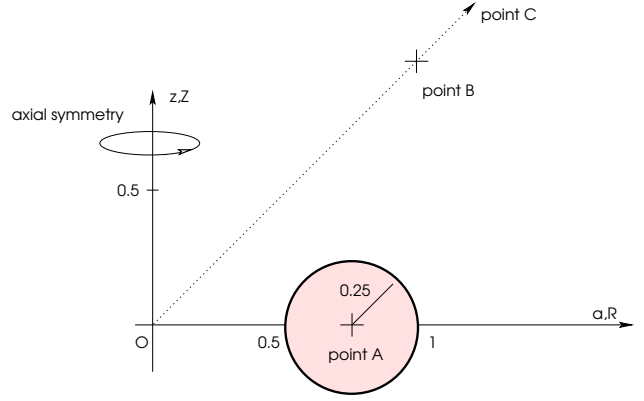}
\caption{A test-torus with circular cross-section, and edges at $\frac{1}{2}$ and $1$ (see Fig. \ref{fig:Tabc.ps} for the mean Green function at points A, B and C).}
\label{fig:torus.eps}
\end{figure}

\section{A numerical example}
\label{sec:numex}

To check the quality of the approximate potential, it is not necessary to
compare Eq.(\ref{eq:psit2}) or Eq.(\ref{eq:psit3}) together with
Eq.(\ref{eq:psit_approx}). It is sufficient to compare the
two Green functions, i.e. $\greenf$ and its approximation $\greenfa$. Since the
approximation made does not involve the $\elih$-function (it is the same for
both expressions), we can simply consider the two members of
Eq.(\ref{eq:intapprox}). To perform this comparison, we have considered a
homogeneous torus with circular cross-section, inner edge $\ain=\frac{1}{2}$
and outer edge $\aout=1$ as depicted in Fig. \ref{fig:torus.eps}. A similar
test-torus has been considered in \cite{ba11}. Also, in order to show how the
mean Green function behaves ($\greenf$ also depends on $R$ and $Z$), we have
selected three points where the potential could be requested:

\begin{itemize}
\item point A$(\frac{3}{4},0)$ inside the torus (at the center),
\item point B$(1,1)$, outside it,
\item point C $(10,10)$, relatively ``far away'' from the system.
\end{itemize}

\begin{figure}
\includegraphics[width=8.5cm, bb=15 46 740 612,clip==]{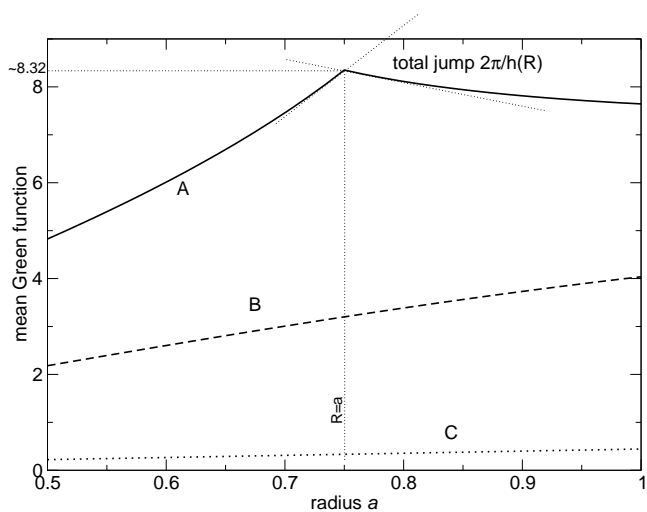}
\caption{Variation  of the mean Green function $\greenf$ with the radius $a$ at points A, B and C for the test-torus considered in Sect. \ref{sec:numex} (see also Fig. \ref{fig:torus.eps}). Inside the system, the radial derivative of the $\greenf$-function undergoes a jump at $R=a$ where the function peaks} (the case of point A), due to the $\elih$-function (see text).
\label{fig:Tabc.ps}
\end{figure}

\begin{figure}
\centering
\includegraphics[width=8.cm, bb=0 0 685 495,clip==, angle=0]{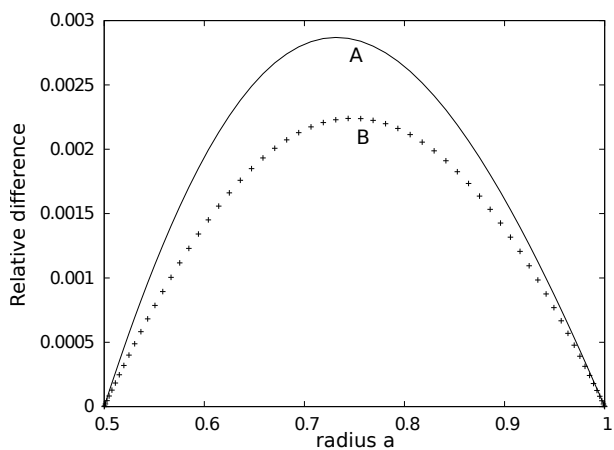}
\caption{Relative difference between the mean Green function $\greenf$ and its approximation $\greenfa$ versus $a$ at points A (plain line) and B (crosses) for the test-torus shown in Fig. \ref{fig:torus.eps}.}
\label{fig:relatAB.ps}
\end{figure}

\begin{figure}
\centering
\includegraphics[width=8.cm, bb=0 0 685 495,clip==,,angle=0]{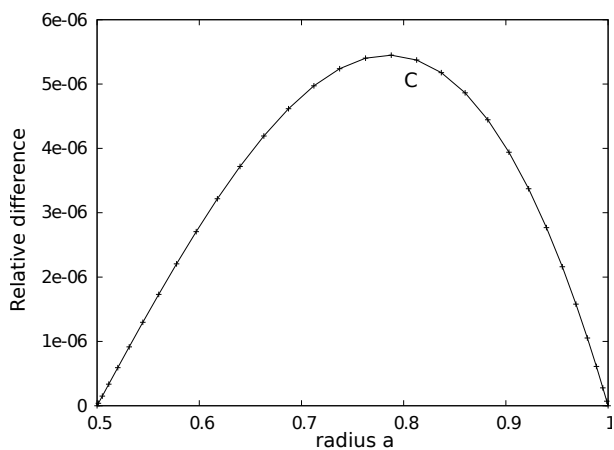}
\caption{Same legend as Fig. \ref{fig:relatAB.ps} but for point C.}
\label{fig:relatC.ps}
\end{figure}

Figure \ref{fig:Tabc.ps} displays the mean Green function $\greenf$ versus the
radius $a$ at points A, B and C. As expected (see Section \ref{sec:theory}),
the function is the largest when $(R,Z)$ stands inside the disc (the case of
point A) and peaks at $a=R$, with a jump in the derivative (due to the
$\elih$-function; see the Appendix \ref{app:c}). At point A, we have $k_\pm
\approx 0.9864$, $\elik(k_\pm) \approx 3.21$ and $\int{k\elie(k)dz} \approx 0.450$ (see i.e. Eq.(\ref{eq:intkdz})), and so $\greenf_{\rm max.} \approx 8.32$ from Eq.(\ref{eq:maxgreenf}). Outside the system (points B and C),
$\greenf$ has a much lower amplitude, but still exhibits a minimum. The
relative difference between $\greenf$ and its approximation $\greenfa$ is shown
in Fig. \ref{fig:relatAB.ps} for points A and B, and in Fig.
\ref{fig:relatC.ps} for point C. According to the Taylor expansion (see Sect.
\ref{sec:approx}), the absolute error is given by $\int_z{(k-\tk)^2dz}$. This
term is easy to calculate. Inside the system and in its close neighbourhood, we find
that the error is $\sim h^4/(a+R)^4$. This is a relatively small error for
geometrically thin discs (most quantities have a precision of $\sim (h/a)^2$ in
general). In the present example where $h$ reaches $0.25$, the absolute error
is expected to be $\sim 10^{-3}$ which is compatible with what we observe for
points A and B. The error in potential values (the area under the curves) is
therefore of the order of  $\sim h^4/(a+R)^4 \times \Delta a$ here, where
$\Delta a$ is the radial extension of the system. This is therefore $\sim 3
\times 10^{-4}$ for point A and for point B, and $\sim3 \times 10^{-6}$ for
point C (see the error map below).

\begin{figure}
\centering
\includegraphics[width=7.4cm, bb=74 130 460 598, clip==, angle=-90]{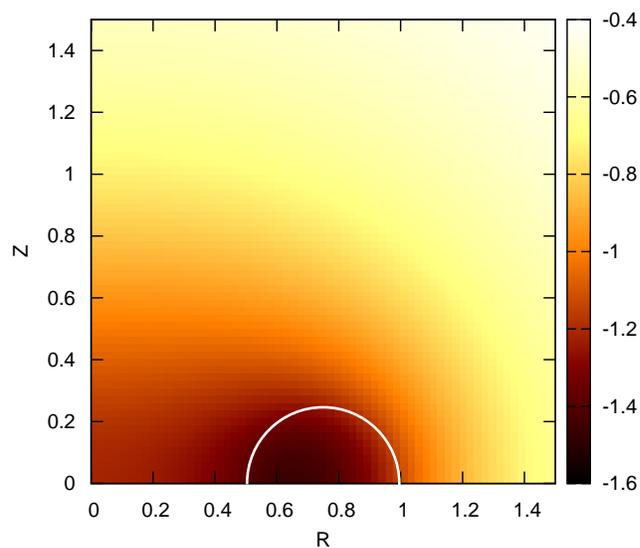}
\caption{Potential of the homogeneous torus computed from Eq.(\ref{eq:psit3}) at the nodes of a regular $(R,Z)$-grid ($64 \times 64$ nodes; $R$ and $Z \in [0,1.5]$). The torus has circular cross-section (shown in white), a diameter $\frac{1}{2}$, and outer edge $\aout=1$.}
\label{fig:psi.ps}
\end{figure}

\begin{figure}
\centering
\includegraphics[width=7.8cm, bb=62 130 460 580, clip==, angle=-90]{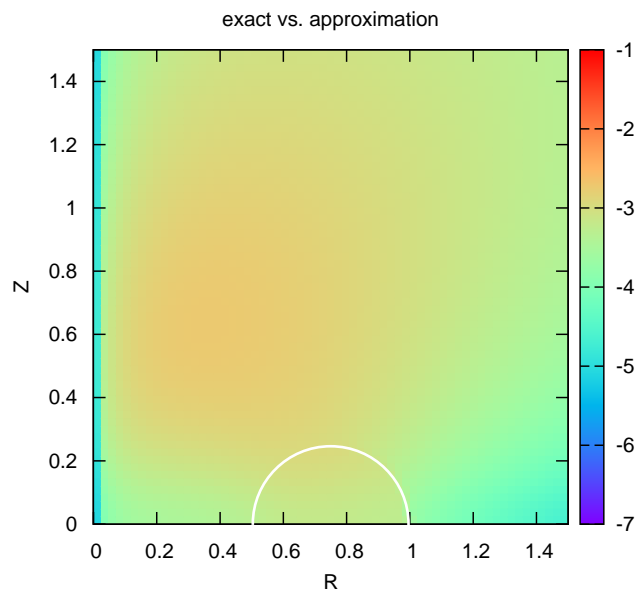}
\caption{Relative error (decimal log. scale) in potential values when using the approximation Eq. (\ref{eq:psit_approx}) rather than Eq.(\ref{eq:psit3}). The conditions are the same as for Fig. \ref{fig:psi.ps}.}
\label{fig:erreur_approx.ps}
\end{figure}

We have computed the potential map in the vicinity of the torus from Eq.
(\ref{eq:psit2}), using the exact mean Green function $\greenf$. The result is
shown in Fig. \ref{fig:psi.ps}. We have also determined the approximate
potential from Eq.(\ref{eq:psit_approx}), that is, using the approximate Green
function $\greenfa$. The relative error (log. scale) between the two potential
maps is shown in Fig. \ref{fig:erreur_approx.ps}. We conclude that we have,
through the approximation, still a very precise estimate of $\int k \elie(k)
d\zeta$ inside as well as outside the system. This is not that surprising since
$\elie(k)$ is a regular function.

 It is worth noting that theactual test-torus, with $\frac{h}{a}~\approx~0.33$, does not correspond to what
is usually called a ``geometrical thin'' system. But as the error is of the
order ${\cal O}(h^4/a^4)$, Eq.(\ref{eq:psit_approx}) remains valid for relatively thick
systems, including thick discs and tori.

\section{Influence of vertical stratification}
\label{sec:inhomogeneous}

\begin{figure}
\includegraphics[width=8.5cm, bb=20 46 740 583, clip==, angle=0]{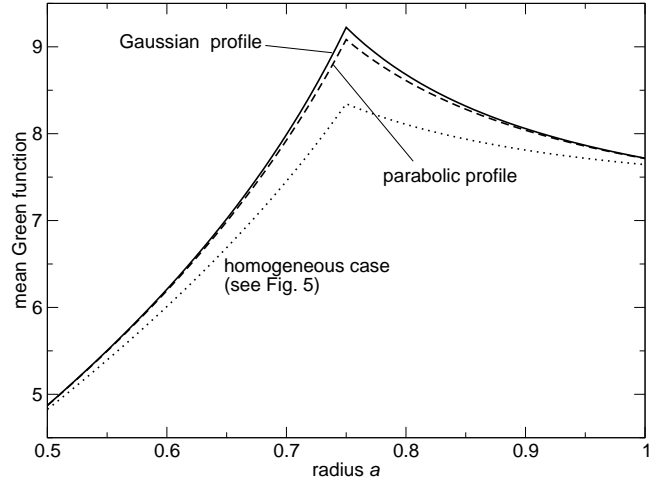}
\caption{Variation of the mean Green function with the radius $a$ at point $A$ for the test-torus considered in Section \ref{sec:numex} (see also Fig. \ref{fig:torus.eps}) for three different  profiles $\rho(z)$ : the homogeneous case (see Fig. \ref{fig:Tabc.ps}), the parabolic profile and the Gaussian profile.}
\label{fig:G_z_rho.eps}
\end{figure}

\begin{figure}
\centering
\includegraphics[width=7.8cm, bb=62 130 460 580, clip==, angle=-90]{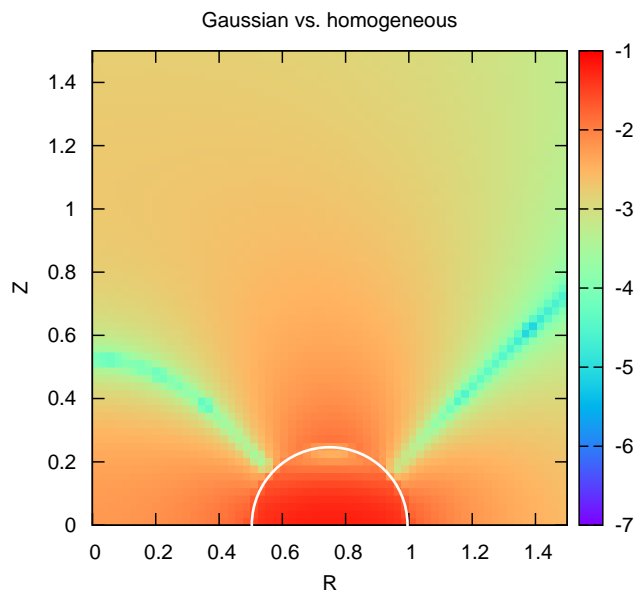}
\caption{Relative error (decimal log. scale) in potential values between the Gaussian case and the homogeneous case (see text for more details). The conditions are the same as for Fig. \ref{fig:psi.ps}.}
\label{fig:gaussianvshomo.eps}
\end{figure}

As realistic systems are expected to have density gradients in the three
directions, it is important to check the sensitivity of the results obtained so far
to the $\rho(z)$ profile. This can easily be done,
still assuming axial symmetry, by introducing the {\it inhomogeneous analog} of Eq.(\ref{eq:meang})
defined by:
\begin{equation}
\label{eq:meanginh}
\greenfinh \equiv \slfrac{\int_{z} 2 \sqrt{\frac{a}{R}} \rho(z) k \elik(k) dz}{\int_{z}{ \rho(z) dz.}}
\end{equation}
For geometrically thin disc models, the mass
density often consists in a power-law of the radius combined with a Gaussian
profile in the direction perpendicular to the midplane, which is a direct
consequence of a locally  isothermal gas in hydrostatic equilibrium
\citep{pringle81,mukm12}. As long as we can decouple the radial and vertical
density structures, an interesting profile is therefore the following
\begin{equation}
\rho(z) = 2 \rho_0 \sqrt{\frac{2}{\pi}} \exp \left( -\frac{2z^2}{h^2}\right),
\end{equation}
where  $z \in ]-\infty,\infty[$ and $\rho_0$ is a constant, possibly a function of the radius. We have also considered the parabolic profile, namely
\begin{equation}
\rho(z) =  \frac{3}{2} \rho_0 \left[ 1 - \left( \frac{z}{h}\right)^2 \right],
\end{equation}
where $z \in [-h,h]$. This profile is very similar to the Gaussian law, matter being however confined into a finite domain. For these two inhomogeneous profiles, the surface density is $2\rho_0 h$, as for the vertically homogeneous case, while central values are different ($1.5$ and $\approx 1.59$ and $1$, respectively). Figure \ref{fig:G_z_rho.eps} shows $\greenfinh(a)$ computed from Eq.(\ref{eq:meanginh}) for the two inhomogeneous profiles for the torus considered in the previous section, at the interior point A. We see that the mean Green functions are very similar in shape, still peaking at $a=R$. The maximum is higher for the parabolic case and even higher for the Gaussian case ($10\%$ and $11 \%$ respectively with respect to the canonical case), due to larger and larger central values\footnote{More generally, for a vertical profile of the form:
\begin{equation}
\rho(z) =  A+B\left( \frac{z}{h}\right)^2,
\end{equation}
we can find a good approximation for $\greenf_{\rm max.}$ by using the right expansion for $\elik(k)$ (see note \ref{note:lnkprim}) followed by integrations terms by terms. We then get:
\begin{flalign}
\greenf_{\rm max.} \approx 2 \Sigma \ln \frac{8R}{h(R)} + 2Ah + \frac{2}{9}Bh
\end{flalign}
which leads to the correct values in the cases considered here, namely: $\sim 8.32$ for the homogeneous profile, $\sim 9.02$ for the parabolic profile and $\sim 9.19$ for the Gaussian profile.}. The difference on potential values (the area under the curves) is however not that large, of the order of $5\%$ (we have $\Psi_A \approx -1.504$ and $-1.513$ respectively, compared to $-1.436$ in the homogeneous case). The deviation is in excess, as expected: in general, the more concentrated the mass distribution, the deeper the potential well. We have performed the same comparison at points B and C which stand outside the torus. We have noticed that as we move away from the distribution, the relative difference between profiles gets smaller and smaller. Again, this is not really a surprise since at large distance, the potential tends to that of a point mass and the details of the distribution are no longer perceptible. Figure \ref{fig:gaussianvshomo.eps} shows the relative difference between the potential generated by a Gaussian profile and the homogeneous case. It turns out that the potential inside as well as outside the body is mainly sensitive to the surface density, which is therefore the critical parameter.

\section{Conclusion}

In this article, we have proposed an alternative definition of the potential of axially symmetrical bodies which avoids a singular kernel. This is
based on the effective integration of the genuine kernel along the vertical
direction. This naturally leaves a regularized kernel, called ``mean Green
function'', which peaks at the location of the initial logarithmic singularity
but remains of finite amplitude. This is yet another proof that the local
contribution of matter located at $\vec{r}' \approx \vec{r}$ is very important
and may dominate the total contribution. This new but equivalent form is
therefore attractive because the corresponding gravitational potential is then
accessible from two classical integrals: the one over the body's meridional
cross-section, and the other over the body's equatorial radius. The total
absence of diverging kernels means that there is no need for special meshing or
specific numerical schemes. Also, the properties of Newton's law are fully
conserved with this approach.

Rigorously, our results are valid only for mass density profiles which are uniform along the
$z$-axis, while there is no restriction about the (radial) surface density profile, thickness/shape and size of the system. This still makes the method relevant to various kinds of configurations, ranging from geometrically thin discs and rings found in different contexts \citep{cd90,dubrulle92,as94}, to possibly long,
cylindrical/filamentary structures as observed in the interstellar medium
\citep{curry00,hennebelle03}. The fact that the asymptotic, long-range behavior of
the potential is well reproduced means that our formula can also be used to
generate quickly and simply grids of potential in which the motion of
test-particles can be studied. Nevertheless, we have analyzed a few
inhomogeneous situations, in particular the Gaussian profile. It turns out that
the sensitivity to stratification is weak on potential values. The local surface density
remains the decisive parameter.

We failed to convert the mean Green function into a line integral | see
Eq.(\ref{eq:keofkdzeta}) and note \ref{not:sech} | as done in \cite{ansorg03} with the genuine kernel
$k\elik(k)$. This does not seem possible unless new closed-form relationships
between complete elliptic integrals with homothetical moduli are derived. Also, we
have expanded at first-order the mean Green function around vanishing aspect
ratio $h/a$, enabling to perform this conversion. We have then obtained an
approximation with very small error. Accuracy can be improved by considering
next terms.\\

\section*{Acknowledgments}
It a pleasure to thank M. Ansorg, A. Bachelot, A. Dieckmann, J. Klinowski for opinions on different aspects of theoretical calculus, and M.-P. Pomies. We wish to thank the referee who suggested to comment about the influence of vertical stratification, which has lead to Section \ref{sec:inhomogeneous}.

\bibliographystyle{mn2e}

\appendix

\section{Definitions}
\label{app:a}

The complete elliptic integral of the first, second and third kinds are defined by \citep{gradryz65}:
\begin{equation} 
\elik(k)=\int_0^{\pi/2}\frac{1}{\sqrt{1-k^2\sin ^2\phi}}d \phi,
\end{equation}
\begin{equation} 
\elie(k) =\int_0^{\pi/2}\sqrt{1-k^2 \sin^2\phi} \, d \phi,
\end{equation}
and,
\begin{equation} 
\elipi(m,k) =\int_0^{\pi/2}\frac{d \phi}{\left(1-m^2\sin^2\phi \right) \sqrt{1-k^2\sin^2\phi}},
\end{equation}
respectively, where $k$ is the modulus and $m$ is the characteristic or
parameter. In the present study, we have $0 \le k \le m \le 1$, $k$ is defined
by Eq.(\ref{eq:kz}), so that $m=k$ for $\zeta=0$. The function $\elid(k)$ is
defined by:

\begin{equation}
k^2 \elid(k)= \elik(k)-\elie(k).
\end{equation}

The demonstration given in Section \ref{sec:integrable_singularity} is based upon the following partial derivatives with respect to the modulus $k$ \citep{gradryz65}:
\begin{equation}
\partial_k \elik(k) = \frac{1}{k}\left[ \frac{\elie(k)}{{k'}^2} -\elik \right],
\end{equation}
\begin{equation}
k \partial_k \elie(k) = \elie(k) -\elik(k), 
\end{equation}
and \citep{durand64}:
\begin{equation}
\partial_k \elipi(m,k) = \frac{k}{m^2-k^2} \left[ \elipi(m,k)-\frac{\elie(k)}{{k'}^2} \right],
\end{equation}
where $k' = \sqrt{1 - k^2}$ is the complementary modulus.

We also have
\begin{equation}
\frac{\zeta}{2\sqrt{aR}} =\pm \frac{\sqrt{m^2-k^2}}{mk}
\label{eq:zetaofmk}
\end{equation}
where sign $+$ stands for $\zeta>0$, and then
\begin{equation}
\zeta \partial_\zeta k = \frac{k}{m^2}\left(k^2-m^2\right).
\label{eq:zetadzetak}
\end{equation}

\section{The function $\elih$}
\label{app:b}

\begin{figure}
\includegraphics[width=8.5cm, clip=]{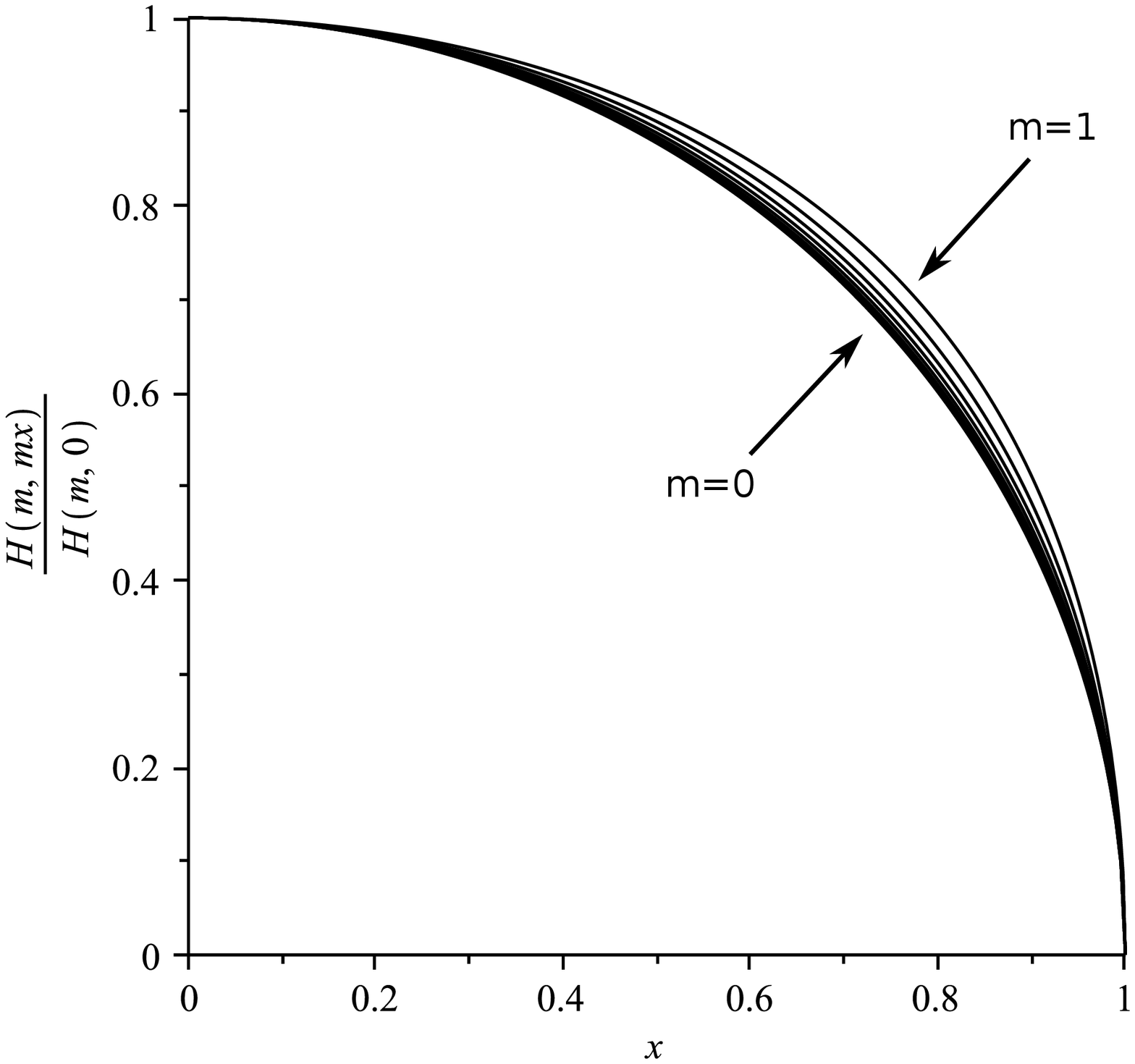}
\caption{The function $\zeta \elih(m,mx)$ normalized to its value at $x=0$, versus $x$ for different values of $m \in [0,1]$.}
\label{fig:zetaH.eps}
\end{figure}

The function $\elih(m,k)$ defined in Section \ref{sec:integrable_singularity} can be rewritten is terms of the variable $x=\frac{k}{m} \in [0,1]$, namely:
\begin{equation}
\elih(m,mx) \equiv mx \left[ \elik(mx) - {m'}^2 \elipi(m,mx) \right],
\label{eq:H}
\end{equation}
and so:
\begin{equation}
\frac{\elih(m,mx)}{\elih(m,0)} =  \frac{2}{\pi} x' \frac{\elik(mx) - {m'}^2 \elipi(m,mx)}{1-m'},
\label{eq:zetaH}
\end{equation}
where
\begin{equation}
\elih(m,0)= \frac{\pi}{2}(1-m'),
\end{equation}
and $x' = \sqrt{1-x^2}$. The ratio $\elih(m,mx)/\elih(m,0)$ is plotted versus $x$ and for different parameters $m$ in Fig. \ref{fig:zetaH.eps}. In a first approximation, we have from the figure $\elih(m,k) \approx x' \elih(m,0)$, and so
\begin{equation}
\zeta \elih(m,k) \approx \pm \frac{\pi}{2} (a+R) (1-m') \frac{{x'}^2}{x}.
\end{equation}

\section{Jump in the radial derivative of the function $\elih(m,k)$}
\label{app:c}

When $a \rightarrow R$ (i.e. $m \rightarrow 1$) and for $\zeta \ne 0$, we have \citep{durand64}:
\begin{equation}
m'^2\elipi(m,k) \underset{m  \rightarrow 1}{\sim} \frac{\pi}{2} \frac{m'}{k'}.
\end{equation} 
As we have 
\begin{equation}
\partial_a m'=\pm \frac{2R}{(a+R)^2}, 
\end{equation}
where sign $+$ stands for $a>R$, we see that the jump in the derivative of $\elipi$ at $a=R$ produces a jump in the derivative of $\elih$. This jump is calculated as follows:
\begin{flalign}
\lim_{m  \rightarrow 1} \partial_a m'^2\elipi(m,k) & =  \partial_a \lim_{m  \rightarrow 1} m'^2\elipi(m,k) \\
\nonumber
& = \partial_a \frac{\pi}{2} \frac{m'}{k'}\\
\nonumber
& = \frac{\pi}{2} \frac{1}{k'}  \partial_a m' - \frac{\pi}{2} \frac{m'}{{k'}^2}  \partial_a\frac{1}{k'}\\
\nonumber
& = \pm \frac{\pi}{2} \frac{1}{k'}  \frac{2R}{(a+R)^2}\\
\nonumber
& = \pm \frac{\pi}{2} \frac{1}{k'}  \frac{1}{2R}
\end{flalign} 
>From Eq.(\ref{eq:zetaofmk}), we have $\lim_{m  \rightarrow 1} \zeta = \pm 2R \frac{k'}{k}$. Consequently, we get for $\zeta <0$:
\begin{equation}
\lim_{m  \rightarrow 1} \zeta k \partial_a {m'}^2\elipi(m,k) =  \pm \frac{\pi}{2},
\end{equation} 
meaning that the jump in the derivative of $\zeta \elih$ amounts to $\pi$ when
$a$ increases and crosses the radius $a=R$. Since the Green kernel contains the
difference $\zeta_+ \elipi(m,k_+) - \zeta_- \elipi(m,k_-)$, the jump is in fact
$2\pi$ in total (see for instance Fig. \ref{fig:Tabc.ps}, point A), except for
a point located on the boundary $(\partial \Omega)$ (i.e. if $Z=z_\pm(a)$),
then it is only $\pi$. This is always true except for  $\zeta=0$. In this case,
the jump disappears and we have:

\begin{equation}
m'\elipi(m,m) \rightarrow \elie(m)
\end{equation} 
which is perfectly continuous.
 
\section{Long-range behavior for the exact potential}
\label{app:d}

At large distance from the body, we have
\begin{equation}
k \sim \frac{2\sqrt{aR}}{r}\rightarrow 0,
\end{equation}
and so the behavior of the elliptic integrals is easily deduced. In particular, we have \citep{gradryz65}:
\begin{eqnarray} 
\begin{cases}
\elik(k) \sim \frac{\pi}{2} \left( 1+\frac{k^2}{4} \right),\\\\
\elie(k) \sim  \frac{\pi}{2} \left( 1-\frac{k^2}{4} \right),\\\\
\elipi(m,k) \sim \frac{\pi}{2} \left[ 1+ \frac{1}{2} \left( m^2 - \frac{k^2}{2} \right) \right].
\end{cases} 
\end{eqnarray}
We then get:
\begin{flalign}
\nonumber
-2G\int{\rho(a) da \sqrt{\frac{a}{R}} \int{k \elie(k) d\zeta} } & \sim  -4 \pi G  \frac{1}{r} \int{\rho(a)  h(a) a da }\\
&  \sim - \frac{GM}{r}
\end{flalign}
which is the expected result. It can be checked that the contribution due to the $\elih$-function is much smaller and behaves like $k^3$. Actually, we find:
\begin{flalign}
\nonumber
\zeta_{\pm}\elih(m,k_{\pm}) & \sim \zeta_{\pm}k_{\pm} \frac{\pi}{2} \left[  \frac{k_{\pm}^2}{2} - \frac{1}{2} \left( m^2 + \frac{k_{\pm}^2}{2} \right)(1-m^2) + m^2 \right] \\
& \sim \zeta_{\pm}k_{\pm} \frac{\pi}{4} \left(1+m^2\right) \left( m^2 + \frac{k_{\pm}^2}{2} \right).
\end{flalign} 
As $k_{\pm}^2$ et $m^2$ are of the same order, $\zeta_\pm\elih(m,k_\pm)$ behaves like $k_{\pm}^3$.

\section{Long-range behavior for the approximate potential}
\label{app:e}

The modulus $k_\pm$ can be approximated as follows:
\begin{equation} 
k_\pm \sim \frac{2\sqrt{aR}}{r}\left(1-\frac{aR+Zz_\pm}{r^2}\right),
\end{equation}
therefore the mean value $\tk$ becomes:
\begin{flalign}
\nonumber
\tk & \sim \frac{2\sqrt{aR}}{r}\left[1-\frac{aR+(z_++z_-)Z}{r^2}\right]\\
& \sim \frac{2\sqrt{aR}}{r}\left(1-\frac{aR}{r^2}\right)
\end{flalign}
We then get for $\elitun$:
\begin{flalign} 
\elitun(\tk)(\zeta_{+}-\zeta_{-}) &  \sim \frac{\pi}{2} h \tk^3\\
&\sim 4\pi h \frac{\sqrt{aR}^3}{r^3} \left(1-\frac{aR}{r^2}\right)^3.
\nonumber
\end{flalign}
Regarding the second term, we have:
\begin{equation} 
\elitdeux(\tk) \sim \frac{\pi}{2}\left(1- \frac{3}{4} \tk^2\right),
\end{equation}
and
\begin{equation}
\int{k d\zeta} \sim k \zeta.
\end{equation}
This order is sufficient. We then find:
\begin{equation} 
\elitdeux(\tk) \int{k d\zeta} \sim - \pi \tk h \left(1- \frac{3}{4} \tk^2\right)
\end{equation}
and we see that this second term dominates over the first one. We then get:
\begin{flalign}
\nonumber
-2G\int{\rho(a) da \sqrt{\frac{a}{R}} \int{k \elie(k) d\zeta} } & \sim  -4 \pi G  \frac{1}{r} \int{\rho(a)  h(a) a da }\\
&  \sim  \frac{GM}{r}
\end{flalign}

\end{document}